\magnification=1095

\input ./macros

\pageno=0

\null

\vskip 2.5truecm

\tolerance=500

\centerline{\vbox to 10truecm{\hsize=11truecm
            \baselineskip=12pt\parindent=0pt\parskip=0pt
            \def\\{\hfil\break}
            \null\vfil
            {\fourteenrm A parallel integration method
             \vadjust{\vrule depth 0.5ex width 0pt}\\
             for solar system dynamics}
             \bigskip
             Prasenjit Saha \vskip1pt
             Mount Stromlo and Siding Spring Observatories,\\
             Australian National University,\\
             Canberra ACT 0200, Australia \vskip1pt
             {\tt saha@mso.anu.edu.au}
	     \medskip
             Joachim Stadel \vskip1pt
             Astronomy Department FM-20,\\
             University of Washington, Seattle, WA 98195-1580 \vskip1pt
             {\tt stadel@astro.washington.edu}
             \medskip
             Scott Tremaine \vskip1pt
             Canadian Institute for Theoretical Astrophysics,\\
             McLennan Labs, University of Toronto,\\
             60 St.~George St., Toronto M5S~3H8,
             Canada \vskip1pt
             {\tt tremaine@cita.utoronto.ca}
             \vfil
}}

\vfil

\noindent {\bf Abstract:} We describe how long-term solar system orbit
integration could be implemented on a parallel computer. The
interesting feature of our algorithm is that each processor is
assigned not to a planet or a pair of planets but to a
time-interval. Thus, the 1st week, 2nd week,\dots, 1000th week of an
orbit are computed concurrently. The problem of matching the input to
the $(n+1)$-st processor with the output of the $n$-th processor can
be solved efficiently by an iterative procedure. Our work is related
to the so-called waveform relaxation methods in the computational
mathematics literature, but is specialized to the Hamiltonian and
nearly integrable nature of solar system orbits.  Simulations on
serial machines suggest that, for the reasonable accuracy requirement
of $1''$ per century, our preliminary parallel algorithm running on a
1000-processor machine would be about 50 times faster than the fastest
available serial algorithm, and we have suggestions for further
improvements in speed.

\vfil

\centerline{Version of $2^{\rm nd}$ May 1996}

\eject

\def\lhead{}
\let\rhead=\lhead

\section 1. Introduction

Planetary orbits have now been studied over times of
$10^8$--$10^{10}$ yr using various techniques, and exploring
progressively longer time scales always seems to reveal interesting
new phenomena.  Laskar (1989, 1990, 1994) used a semi-analytic secular
perturbation theory to follow planetary orbits for up to 10 Gyr, and
found weak chaos in the orbits of the inner planets, which led for
example to significant evolution in Mercury's orbit over the lifetime
of the solar system. Sussman \& Wisdom (1992) directly integrated the
orbits of the nine planets for 100~Myr, verifying Laskar's discovery
that the orbits of the inner planets are chaotic, and also finding
weak chaos in the orbits of the outer planets which did not appear in
Laskar's secular theory. Laskar \& Roboutel (1993)  and Touma \&
Wisdom (1993) by different methods found that Mars's obliquity is
chaotic, whereas the Earth's is apparently stabilized by the
Moon. Naturally, long integrations are not ephemerides but probes of
qualitative and statistical properties of the orbits. We remark that
it is possible to extract information about dependence on initial
conditions from Fourier analysis of a single integration
(Laskar~1993).

Since the longest direct orbit integrations are still much shorter
than the age of the solar system, there is motivation to invent better
numerical integration algorithms. The best long-term accuracy is
achieved by methods that exploit (a)~the Hamiltonian, and (b)~the
nearly Keplerian properties of the equations. Such methods were
introduced by Wisdom \& Holman (1991; hereafter \WH) and independently
by Kinoshita, Yoshida \& Nakai (1991), with subsequent improvements by
various authors; for brevity, we will call them `generalized leapfrog'
in this paper, because of their formal resemblance to leapfrog. An
$n$-th order generalized leapfrog integrator can have global error of
O$(\epsilon^2\tau^nT)$, where $\epsilon\leqsim 10^{-3}$ is the
relative scale of planetary perturbations, $\tau$ is the timestep, and
$T$ the total length of an integration.  The $\epsilon^2$ factor and
the linear dependence on $T$ (rather than quadratic, as for general
purpose integrators) make low-order schemes worthwhile. The state of
the art demands an accuracy of $\sim10^{-9}$ in the frequencies (or
$\sim1''$ per century for Mercury), which is
achievable by second-order generalized leapfrog schemes with
$\tau\simeq1\,\rm week$---see Wisdom, Holman \& Touma (1994) or Saha
\& Tremaine (1994; hereafter \ST). But a 5~Gyr integration 
remains prohibitively expensive at current floating-point speeds.

This situation prompts one to think about parallel integration methods, and in
this paper we develop a parallel integrator.  We adopt a very different sort
of parallelism from galactic or cosmological parallel codes; instead of
distributing the force calculation among different processors, we distribute
different time slices among different processors.  Solar system dynamics
invites such `timelike' rather than the usual `spacelike' parallelism because
orbits evolve very slowly compared to orbital times, whereas the number of
forces that could be distributed between processors is relatively small.  The
new algorithm is very similar to second-order generalized leapfrog in
accuracy, and is better at controlling roundoff error; but it is
arithmetically more cumbersome, so---at least in the present version of a
parallel integrator---the speed gain possible is smaller than the number of
processors.

\section 2. A parallel algorithm

Consider some system of ordinary differential equations $\dot z = f(z)$,
or equivalently, a set of quadratures:
$$ z(t) = z(0) + \int_0^t f\left(z(t')\right)\, dt'.  \putnum$$
\name\quadra
We can approximate the quadratures by sums; discretizing time with
timestep $\tau$ and writing $z_n$ to mean $z(n\tau)$, we can write,
for example: 
$$ z_n = z_0 + \tau \sum_{m=0}^{n-1} f\zmid{m}{m+1};
\qquad n=1\,.\,.N. \putnum$$
\name\scan 
This can now be regarded as a set of nonlinear equations in many
variables ($N$ times the dimensionality of $z$), which can be solved
by some iterative process.  In an $N$-processor computer, the right
side of equation (\scan) can be evaluated in parallel, with each
processor computing one term and O$(\log N)$ sums. If the number of
iterations is much less than the number of processors, then
parallelism will yield a speed gain.

Solving the huge set of equations (\scan) in $\ll N$ iterations may
seem a hopeless task. But in fact it can often be accomplished
(provided the original differential equations are very smooth and
there is a good starting guess for the $z_n$) by Newton-Raphson or
variants of it not requiring derivatives (see e.g., Ortega \& Rheinbolt
1970). The general idea is known as {\it waveform relaxation;} we
suggest Bellen \& Zennaro (1989) as an introduction to the literature
in this area.  In this paper, however, we will proceed from first
principles, because for nearly integrable Hamiltonian systems the
solution becomes quite simple.

For a Hamiltonian system with canonical variables $z\equiv(p,q)$ the
particular quadrature formula in equation (\scan) amounts to integrating the
differential equations with the `implicit midpoint' integrator 
$$ \eqalign{ &p_1 = p_0 - \tau\pderiv(/q)H\zmid01, \cr &q_1 = q_0 +
\tau\pderiv(/p)H\zmid01, \cr} \putnum$$\name\fengform 
with timestep $\tau$ (which can be kept fixed when integrating
planetary orbits). This is a second order symplectic integrator, which
is to say that the transformation $z_0\rightarrow z_1$ differs from
the true time evolution by $O(\tau^3)$ but is exactly canonical.  It
is one of a class of symplectic integrators derived by Feng (1986). In
the Appendix we derive a somewhat stronger statement: the integrator
exactly follows the dynamics of a `surrogate' Hamiltonian
$$ ~H = H + H_{\rm err}, \putnum$$\name\Hsurr 
which differs from the exact Hamiltonian $H$ by an error Hamiltonian $H_{\rm
err}$ which is O$(\tau^2)$ 
and non-autonomous ($H_{\rm err}$ is periodic in
time with period $\tau$).  In any case, the symplectic property guarantees
that spurious dissipation will not occur.

Consider now an $H$ that is close to some integrable Hamiltonian, with $p,q$
being action-angle variables for the latter, thus:
$$ H = \Hzer(p) + \epsilon\Hone(p,q,t). \putnum$$\name\sahag
The formula (\scan) for such a system becomes
$$ \eqalign{ &p_n = p_0 - \tau\sum_{m=0}^{n-1}
                  \epsilon\pderiv(/q)\Hone\zmid{m}{m+1}, \cr
             &q_n = q_0 + \tau\sum_{m=0}^{n-1} \left[
                  \pderiv(/p)\Hzer\zpmid{m}{m+1} +
                  \epsilon\pderiv(/p)\Hone\zmid{m}{m+1}
             \right]. \cr}  \putnum$$\name\nstep
The error Hamiltonian is $\hbox{O}(\epsilon\tau^2)$ (see
Appendix). The error has the same order as in second-order generalized
leapfrog, suggesting that the performance will be similar. Generalized
leapfrog is preferable for serial integrations because it is explicit, but
as part of an iterative parallel scheme explicit methods offer no advantage,
and implicit midpoint leads to a simpler formulation.

We now need to solve the equations (\nstep) for the $p_n,q_n$. Let us
abbreviate the equations as
$$ \eqalign{ &p_n - p_0 - u(P_n,Q_n) = 0, \cr
             &q_n - q_0 - V(P_n) - v(P_n,Q_n) = 0,\cr}  \putnum$$\name\sstt
where $u$ and $v$ are O$(\epsilon)$ smaller than $V$, and $P_n$, $Q_n$ are the
vectors $(p_0.\,.p_n)$, $(q_0.\,.q_n)$. Suppose that $P_n,Q_n$ are
currently at some good guess values for the exact solutions $P_n',Q_n'$. We
start out as if to derive a Newton-Raphson iteration, and expand (\prevenum1)
to leading order in $z_n-z_n'$:
$$\eqalign{ &p_n-p_0-u(P_n,Q_n) \simeq [p_n'-p_0-u(P_n',Q_n')]+p_n - p_n' \cr
&\quad - \<\hbox{terms involving derivatives of $u$}> \cr
\noalign{\smallskip}
&q_n-q_0-V(P_n)-v(P_n,Q_n)\simeq
[q_n'-q_0-V(P_n')-v(P_n',Q_n')]+q_n-q_n' \cr &\quad
  \textstyle -\sum_{m=1}^n(p_m-p_m')\pderiv(/p_m)V(P_n)
- \<\hbox{terms involving derivatives of $v$}> \cr}
\putnum$$
The terms in square brackets are zero since $z'$ solves (\sstt).   
We then substitute $V(P_n)-V(P_n')$ for 
$\sum_{m=1}^n(p_m-p_m')\pderivp(/p_m)V(P_n)$ and
neglect derivatives of $u$ and $v$ (thus departing from Newton-Raphson and
sacrificing quadratic convergence), to obtain 
$$\eqalign{ &p_n'\leftarrow p_0 + u(P_n,Q_n), \cr 
&q_n' \leftarrow q_0 + V(p_l') + v(P_n,Q_n), \cr}
\putnum$$\name\iterform 

Once in the neighborhood of the correct solution, the simple iteration
(\prevenum1) will give linear convergence by a factor O$(\epsilon)$ per
iteration. The obvious starting guess to use is the solution to the
unperturbed equations, which is trivial since $p,q$ are action-angles Better
guesses are possible, such as the mean semi-major axes and the eccentricities
and inclinations given by secular perturbation theory; however, our tests with
toy problems suggest that these will not lead to dramatic reductions in the
number of iterations required for convergence. Note that making $N$
excessively large will not introduce the danger of divergence of iterations,
because the sequence of approximants to $z_n$ does not influence the
approximants at any $z_{m<n}$. Provided the iterative process converges for
the serial formula (\fengform), the worst that can happen in the parallel case
is that the convergence becomes proportional to $N$.

\section 3. An simple example

It is helpful to apply the iterative procedure of the previous section as
applied to a simple example. We consider the pendulum Hamiltonian
$$ H = \half p^2 - \epsilon \cos q.   \putnum$$\name\Hpend
For this case, because integrals of the type (\quadra) can be done exactly,
we will leave them as integrals rather than replacing them by sums.  We also
make one change of notation for this section: $p_k$ and $q_k$ will
refer to the $k$-th approximation to $p(t)$ and $q(t)$, and not to values
on a grid.

With these changes the iterative scheme (\iterform) becomes
$$\eqalign{p_0(t) &= p(0),\cr
           q_0(t) &= q(0)+ p(0)t,\cr 
       p_{k+1}(t) &= p(0)-\epsilon\int_0^t\sin[q_k(t')]\,dt',\cr
       q_{k+1}(t) &= q(0)+\int_0^tp_{k+1}(t')\,dt'.\cr}\putnum$$
\name\ierex
We can derive expressions for the errors in the $n$-th approximation
$$  \delta p_k(t)\equiv p_k(t)-p(t),\qquad 
    \delta q_k(t)\equiv q_k(t)-q(t),\putnum$$
by comparing the iterates (\ierex) with the exact equations
$$\eqalign{p(t) &= p(0)-\epsilon\int_0^t\sin[q(t')]\,dt',\cr
           q(t) &= q(0)+\int_0^tp(t')\,dt'.\cr}\putnum$$
Assuming $|\delta q_k|\ll 1$ and hence $\sin[q_k(t)]\simeq\sin[q(t)]$ for
all $k\ge0$, we get:
$$\eqalign{
    \delta p_0(t) &= \epsilon\int_0^t\sin[q_0(t')]\,dt',\cr
    \delta q_0(t) &= \int_0^t\delta p_0(t')\,dt',\cr 
    \delta p_{k+1}(t) &= -\epsilon\int_0^t\cos[q_0(t')]\,
                         \delta q_k(t')\,dt',\cr
    \delta q_{k+1}(t) &= \int_0^t\delta p_{k+1}(t')\,dt'.\cr
  }\putnum$$
These equations can be solved analytically, although the expressions for
$\delta p_k(t)$ and $\delta q_k(t)$ rapidly become quite complicated as $k$
grows. For large values of $\epsilon t$ we find that to leading order the even
iterates satisfy 
$$ |\delta q_{2k}(t)| = {|\cos q(0)|\over 2^k\,(2k+1)!}
       \left({\epsilon t\over p(0)}\right)^{2k+1}  + \hbox{O}(t^{2k-1});
       \qquad |\delta p_{2k}(t)| = {(2k+1)\over t}|\delta q_{2k}(t)| +
       \hbox{O}(t^{2k-1}).    \putnum$$
Thus the iterative scheme will converge for all $\epsilon t$. For
small $\epsilon t$ the number of iterations required will be roughly
$n_{\rm it}\propto-1/\ln(\epsilon t)$. For $\epsilon t\gg1$, we expect
that if $p(0)=1$, then $n_{\rm it}\propto \epsilon t$ iterations will
be required for convergence. 

\fig[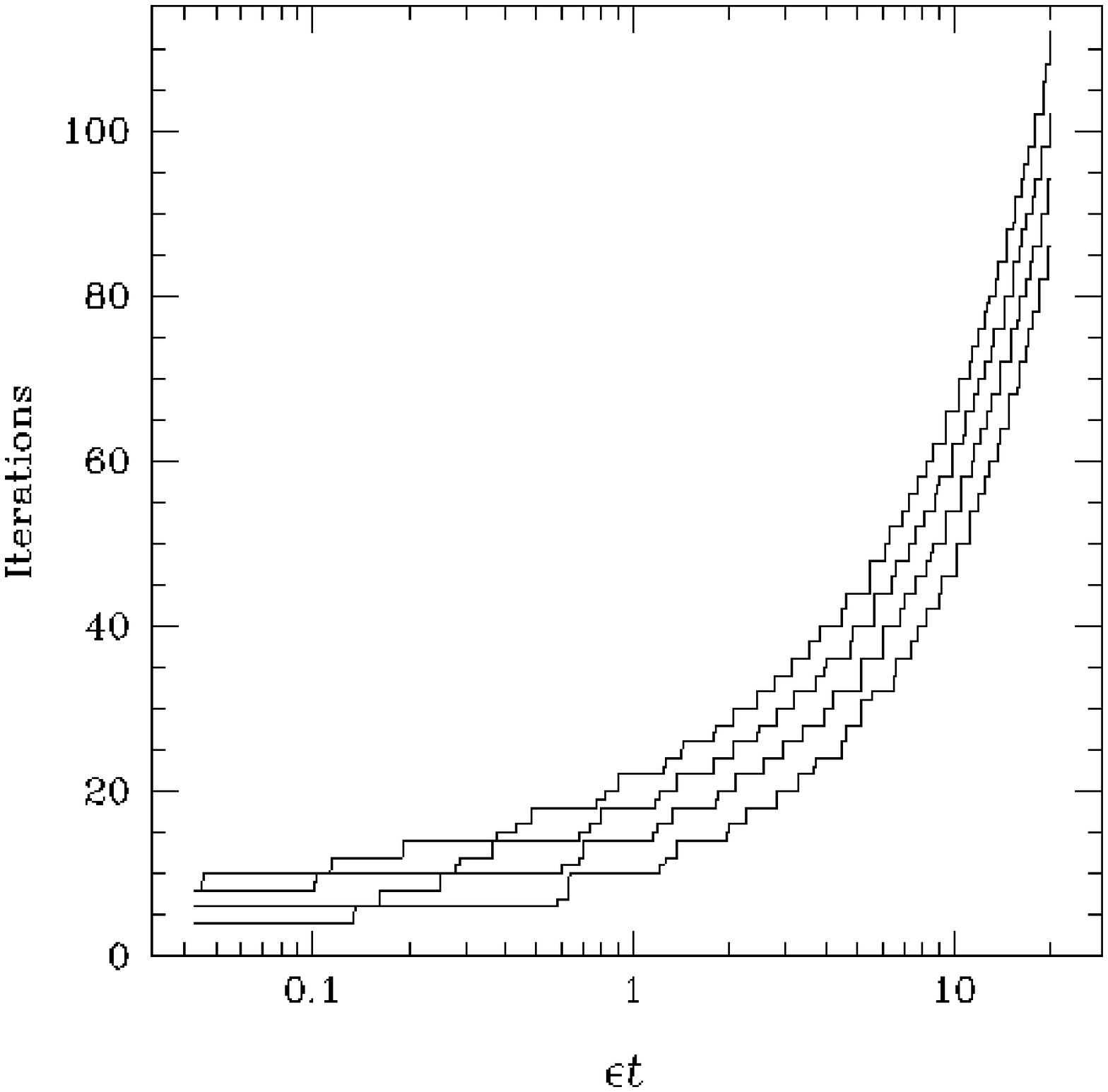,\figsize,\figsize] Figure 1. Diagnostics from a numerical
solution for the pendulum Hamiltonian (\Hpend) with $\epsilon=0.01$,
$p(0)=1,q(0)=0,$ integrated using the algorithm of Section 2, with
timestep $\tau=0.1$.  The four curves are the number of iterations
required for convergence to $10^{-4}$, $10^{-6}$, $10^{-8}$, and
$10^{-10}$. The corresponding plot for $\tau=0.01$ is almost
identical.

We plot in Figure 1 the number of iterations taken by the scheme
(\iterform) to converge to various levels as a function of time. This
case differs in detail from the one we have just analyzed, because it
refers to a discretized situation as in Section~2; this means that the
iterations are converging not to the true solution, but to a surrogate
differing from the true solution by a timestep-dependent amount. Also,
the approximation $\sin[q_k(t)]\simeq\sin[q(t)]$ we made above is not
always valid. Nevertheless, Figure~1 agrees with our qualitative
prediction of number of iterations $\propto\epsilon t$ for large
$\epsilon t$; quantitatively, the number of iterations is $\approx
4\epsilon t$.

\section 4. Surrogates, and warmup

If a generalized leapfrog integrator is applied to a nearly integrable
Hamiltonian, the output can be expressed formally as the time
evolution under the surrogate Hamiltonian
$$ \widetilde H = H + \epsilon\tau^n\times\<\hbox{power series in $\epsilon$
            and $\tau^2$}>, \putnum$$\name\surrser
for an $n$-th order scheme.  We say `formally' because the series in equation
(\prevenum1) does not in general converge; nevertheless, analysis of this
series is useful.  In particular, the $\epsilon\tau^n$ indicates long term
errors of O$(\epsilon\tau^nT)$. It turns out that one can eliminate the effect
of secular O$(\epsilon)$ terms in $\widetilde H$ by applying a special small
transformation to either the initial conditions or the output, thus reducing
the errors to O$(\epsilon^2\tau^nT)$.  Various transformations which
accomplish this are given in Saha \& Tremaine (1992), \WHT, and \ST. Probably
the simplest comes from the latter paper, and consists of the following
`warmup' to the main integration.
\pseudocode
     \<Integrate backward (or forward) for a large number of orbits
       with the timestep $\tau$ scaled down by
       $\leqsim\epsilon^{1/n}$ while linearly reducing the
       perturbation strength to zero\>
     \<Integrate forward (or backward) to the starting point with the
       regular timesteps $\tau$, while linearly reviving the
       perturbation strength to the correct value\>
\endcode
In the Appendix, we show that errors in the implicit midpoint integrator also
can be described by a surrogate Hamiltonian, which differs
from $H$ by an error Hamiltonian that is O$(\epsilon\tau^2)$ if $H$ is nearly
integrable. Thus the above warmup procedure will reduce long-term errors from
O$(\epsilon\tau^2T)$ to O$(\epsilon^2\tau^2T)$ 
whether we use generalized leapfrog or implicit midpoint. In fact, the results
from the implicit midpoint method, both without and with warmup, are very
similar to those from the corresponding generalized leapfrog method.

\section 5. A simulation with nine planets

To apply the new integration method to planetary orbits, we must:
\item{(i)} reduce the Hamiltonian to the form $\Hzer+\epsilon\Hone$,
where now $\Hzer$ is Keplerian and $\epsilon$ represents the smallness of
planetary perturbations;
\item{(ii)} find action-angle variables $p,q$ for $\Hzer$;
\item{(iii)} find expressions for $\pderivp(\Hone/p)$ and
$\pderivp(\Hone/q)$.

Item (i) is standard, though not trivial. Barycentric variables will
not do because the Sun's motion is not nearly Keplerian. The key (see
WH91) is to take Mercury's coordinates relative to the Sun, Venus's
relative to the barycenter of Sun-Mercury, and so on (the ordering of
planets is immaterial). The time derivatives of such
coordinates---called the Jacobi coordinates---turn out to be
proportional to the conjugate momenta. The Sun's motion drops out and
is replaced by the ignorable motion of the solar system
barycenter. The Hamiltonian has the desired form, with one Keplerian
term for each planet plus perturbation terms of $O(\epsilon)$.
Post-Newtonian effects from general relativity add a further
complication; a procedure for handling these effects is described in
\ST.

For item (ii), we transform the Jacobi variables for each planet to
Poincar\'e elements $\Lambda,\lambda$, $\xi_1,\xi_2$,
$\eta_1,\eta_2$. These relate to the more familiar Keplerian elements
$a$, $e$, $i$, $\Omega$, $\omega$, $M$ as follows:
$$ \eqalign{
     &\Lambda = \sqrt{\mu a}, \quad \lambda=\Omega+\omega+M, \cr
     &\xi_1 =  \left[2\Lambda(1-\sqrt{1-e^2})\right]^\half \cos(\Omega+\omega)
      \simeq\sqrt\Lambda\,e\cos(\Omega+\omega), \cr
     &\xi_2 = -\left[2\Lambda(1-\sqrt{1-e^2})\right]^\half \sin(\Omega+\omega)
      \simeq -\sqrt\Lambda\,e\sin(\Omega+\omega), \cr
     &\eta_1 =  \left[2\Lambda\sqrt{1-e^2}(1-\cos i)\right]^\half \cos\Omega
      \simeq\sqrt\Lambda\,i\cos\Omega, \cr
     &\eta_2 = -\left[2\Lambda\sqrt{1-e^2}(1-\cos i)\right]^\half \sin\Omega
      \simeq-\sqrt\Lambda\,i\sin\Omega.\cr
  }  \putnum$$
The first two are action-angles of $\Hzer$; the other four are not
action-angles but they are canonical, which is good enough since
$\Hzer$ involves only the $\Lambda$ of each planet. Poincar\'e
elements remain non-singular for zero $e$ or $i$.

Item (iii) consists of transforming the accelerations to Poincar\'e
variables. This is straightforward with the chain rule, but boring to
program and expensive to compute. With nine planets, we found that
transforming accelerations from cartesian Jacobi variables to
Poincar\'e elements took more than four times as long as evaluating
the accelerations in the first place.

\fig[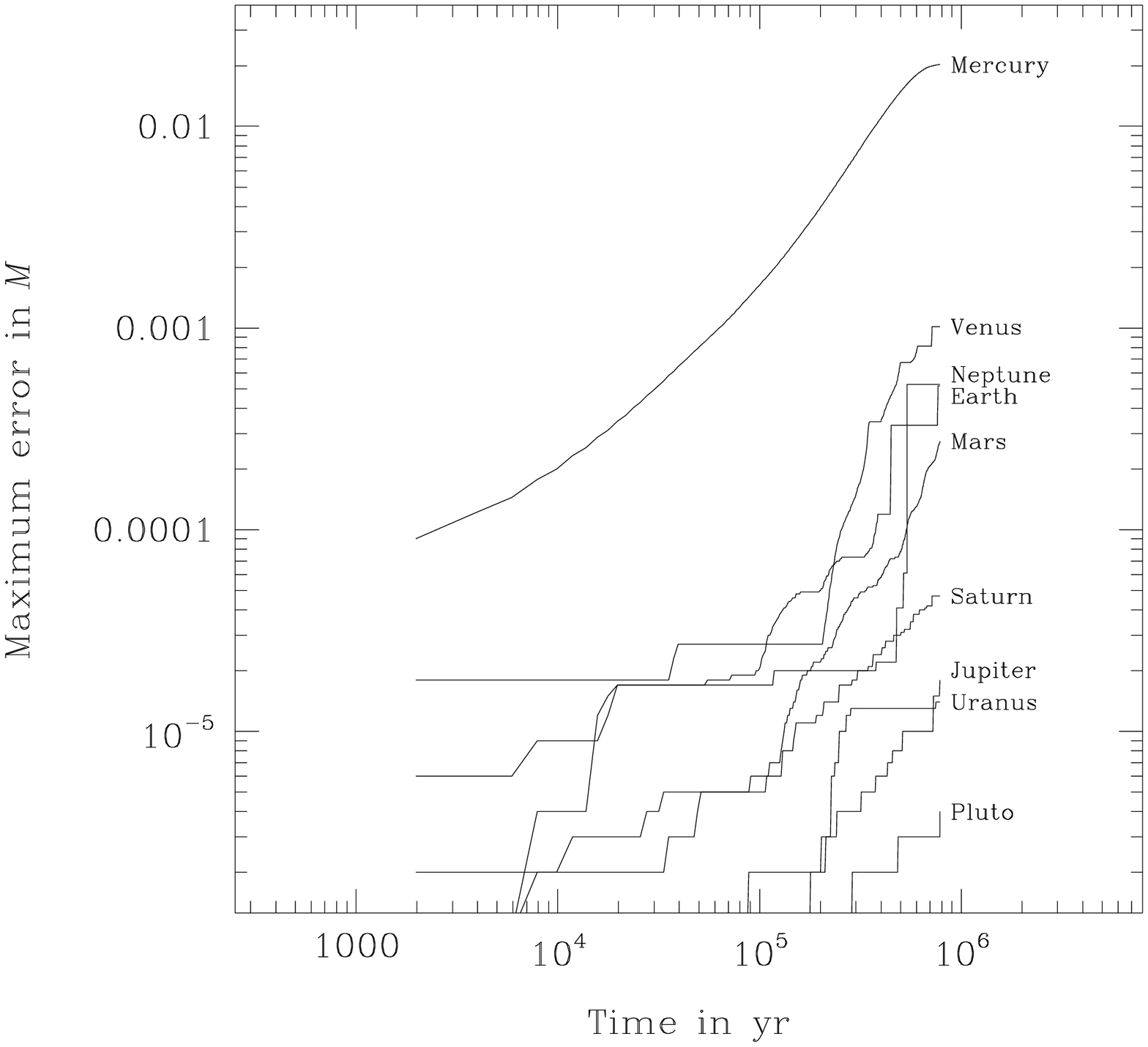,\figsize,\figsize] Figure 2. Maximum error in mean
anomaly $M$ up to time $T$, against $T$, with a timestep of
$7\frac1/{32}$ days.

We have programmed the algorithm to simulate its operation on a parallel
computer, and tested the results of a medium-length integration.  We
used a timestep of $7\frac1/{32}\,\rm days$ and a warmup length of
5000~yr with timestep scaled down by 32.  Our single actual processor
was simulating 4096 concurrent ones. Figure~2 shows the estimated errors
in our integration, gotten by comparing with a more accurate version of
the integration by Quinn, Tremaine \& Duncan (1991). It is quite similar
to Figure~2 in \ST\ which shows error estimates for generalized leapfrog
with the same timestep.

\fig[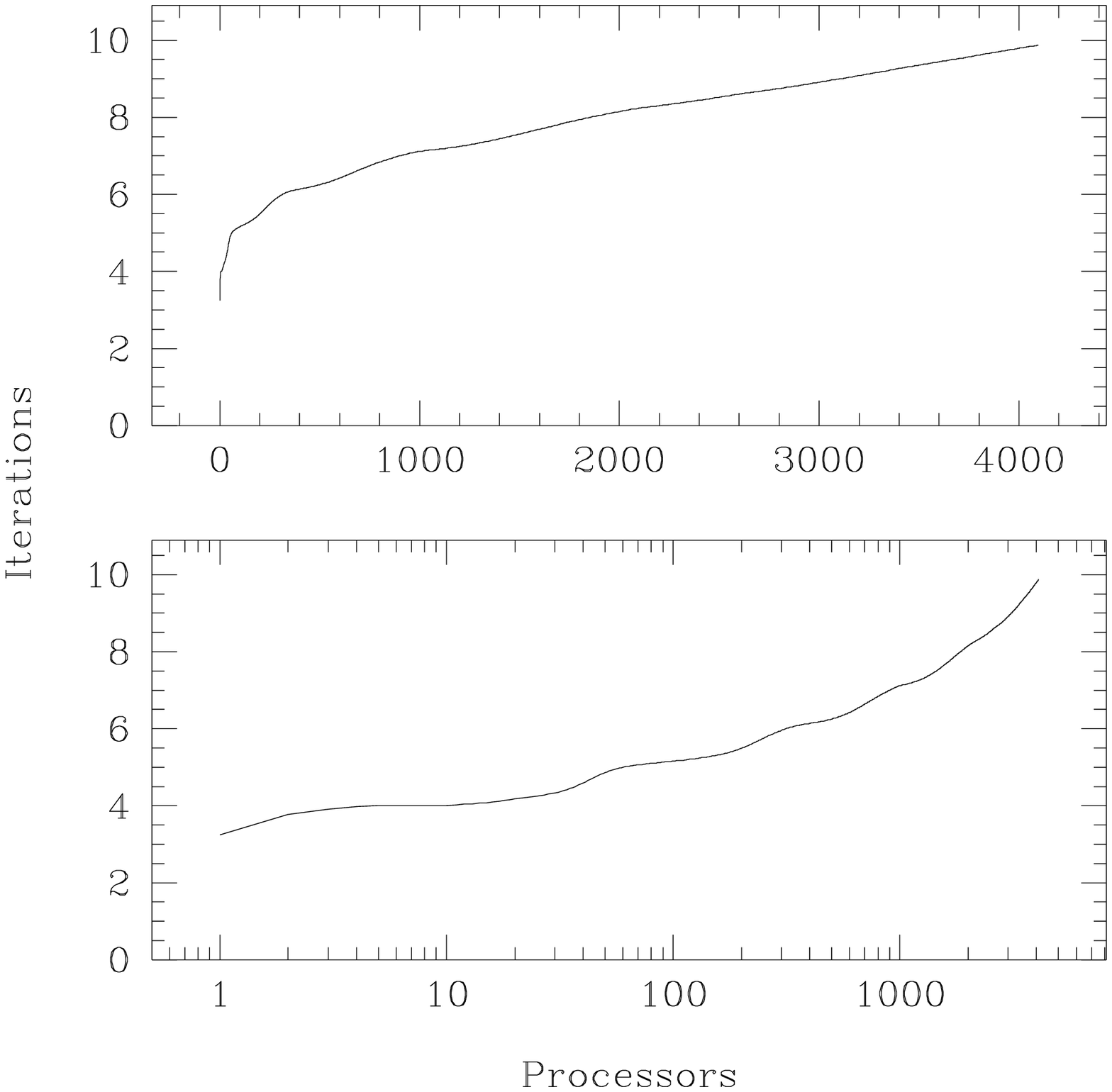,\figsize,\figsize] Figure 3. The average number of iterations
required for convergence (to $\simeq50$ bits of precision) versus the
number $N$ of simulated processors.  The two panels show the same data
using linear and logarithmic scales for $N$.

In Figure~3 we show the average number of iterations $n_{\rm it}$ our algorithm
needed to converge to $\simeq50$ bits of precision (relative error
$\simeq10^{-15}$), over blocks of steps of different lengths $N$ (in effect,
different numbers of processors). The plots exhibit the linear dependence at
large $N$ predicted by the simple model of Section 3, and can be fitted by
$$n_{\rm it}\simeq 6+\left(N\over 1000\right)\qquad\hbox{for } N>
500.\putnum$$
\name\iterpred

While the above results demonstrate that our algorithm works, our current
implementation is probably far from optimal. In particular, our program does
about 10 times as much arithmetic per iteration as generalized leapfrog with
individual timesteps to complete an integration of the same length. The
picture changes, though, when we consider the control of roundoff errors. We
consider this and other prospects for improvement in the concluding section.

\section 6. Prospects

A serious obstacle to accurate long solar system integrations is roundoff
error (e.g.\ Quinn \& Tremaine 1990). Roundoff errors generally act like
spurious dissipation, introducing secular drift in the energy, and hence
positional errors $\propto T^2$. More precisely, if $\delta$ is the floating
point precision, and $P$ the orbital period, the expected longitude error
after time $T$ is $\Delta\lambda\approx T^2\delta/(\tau P)$. For a 5~Gyr
integration with a 1~week timestep, at least 76 bits are needed to keep the
roundoff contribute to Mercury's longitude error below 0.01
radians. (Of course,
0.01 radians is much smaller than the expected integration errors, which are
inflated by weak chaos, but it is good to set low tolerances for roundoff
error, because it is an unphysical effect.)  IEEE double precision provides a
mantissa of only 53 bits. Many compilers offer quadruple precision, but that
is implemented in software, and is slower by a factor of 20 at least.

Our new method offers a significant improvement in effective precision with
only a small sacrifice of speed. The reason is that the perturbative impulses
($u$ and $v$ in equation [\iterform]) are by far the major computational
cost. For a 1~week timestep and $\epsilon\sim10^{-3}$, these impulses are
$10^{-4}$ smaller than $p,q$; hence they require $\simeq13$ bits less
precision.  So by doing the force calculation in double precision and the few
remaining operations in quadruple precision, one can effectively gain
$\simeq13$ bits of precision. 

Also, some floating point units (including the Intel $n86$ series and the
Motorola $68n$ series) use 80 bits internally and compilers for these may
offer an extended double precision type with a 64-bit mantissa. On such
hardware, we expect that an effective precision of $\simeq77$ bits will cost 
$<20\%$ more time than IEEE double precision. 

Similar games can be played with other integrators, of course, but
generally with less success. In multistep integrators (see Quinlan \&
Tremaine 1990) force includes the solar part, so the gain in effective
precision is less. In generalized leapfrog, there is a large subset of
operations not involving small quantities, and when done in quadruple
precision these slow the method $\simeq6$ fold. 

In summary, when roundoff is considered, we estimate that the parallel
algorithm running on 1000 processors will be about 50 times faster than
generalized leapfrog (as implemented in \ST) on a single
processor. 

To close this paper, we mention some avenues which may lead to
more efficient parallel algorithms. 

\item{1)} Rather than devote 1 processor to each of $N$ timesteps, it might
be better to set $K$ processors working in parallel on each of $N/K$
timesteps. Let us assume (i)~that using $K$ processors reduces the
time needed to complete a timestep by a factor $K/g$ where $g>1$
represents the degree of parallelism ($g$ will generally be an
increasing function of $K$), and (ii)~that the number of iterations
required for convergence at timestep $m$ is $n(m)=cm+n_0$, where $c$
and $n_0$ are constants (cf.\ eq.\ \iterpred). Then parallelizing each
timestep is faster if the number of processors $N>n_0K(g-1)/[c(K-g)]$;
for $K\gg1$ and the parameters of equation (\iterpred) this yields
$N>6000(g-1)$.

\item{2)} The iterations converge much more quickly near the start of
a block of timesteps than near the end. Thus the early processors
become free while the later ones are still converging. These freed
processors could immediately start working on the next block of
timesteps, using the partial converged results from
the previous block as a starting point.

\item{3)} In our implementation, we have not paid much attention to
algebraic simplifications, using the chain rule to convert accelerations from
cartesian variables to Poincar\'e elements.  It would be worth some effort to
try to recast the change of variables, for the chain rule is an arithmetic
glutton.  An intriguing possibility is that one might be able to dispense with
the Poincar\'e variables, and simply use the method on the variables $a$,
$\lambda$, $e\cos(\Omega+\omega)$, $e\sin(\Omega+\omega)$, $i\cos\Omega$,
$i\sin\Omega$; the accelerations would be just the right sides of Gauss's
equations for these variables, which are relatively simple to evaluate.  Now,
the latter variables are not canonical, so the integrator will not be
symplectic; but it will still be time-symmetric, and possibly that is good
enough. An alternative approach would be to design an iteration scheme based
on a generalized leapfrog integrator rather than the implicit midpoint method.

\item{4)} Our iteration scheme (\iterform) is only linearly convergent. It
would be interesting to seek an efficient quadratically convergent iteration
scheme. 

\bigskip

We thank Wayne Hayes, Matt Holman, Agris Kalnajs, and George Lake for
discussions. PS is grateful to the University of Washington for their
hospitality. This research was supported by NSERC.

\def\thenumber{{\rm A}\the\eqnum} \eqnum=0

\sectionlike Appendix: The implicit midpoint method

This Appendix derives the error Hamiltonian associated with the implicit
midpoint method (\fengform), following Feng (1986).

Let $\z$ and $\w$ be vectors of the type $(p_1.\,.p_K,q_1.\,.q_K)$ in
phase space.  In the following we will use matrix notation: $\z$
denotes a column vector, $\z^T$ the corresponding row vector, and so
on; $\J$ is the usual symplectic matrix.

Consider a map that takes $\z(0)$ to $\z(t)$, through a generating
function. We write
$$ \eqalign{ &\z = \z(t), \quad \z_0 = \z(0), \cr
             &\w = \half(\z+\z_0), \quad \phi = \phi(\w,t), \cr
   } \putnum$$
and subscripts for gradients and time derivatives, like $\phi_\w$ and
$\phi_t$. The map is
$$ \z - \z_0 = \Jinv\phi_\w, \putnum$$\name\genfunc
($\phi_\w$ being a gradient).  This map is canonical.\note{There are

several ways of proving that maps of the type (\genfunc) are
canonical, and here is one. 
If $\x$ and $\y$ are two phase space
vectors, then the map $\x\to\y$ will be canonical if $\x^T\J\,d\x -
\y^T\J\,d\y$ is an exact differential. Let $\w=\half(\x+\y)$, and suppose
that
$$ \half(\x^T\J\,d\x - y^T\J\,d\y) = d\phi + d(\w^T\J\x),  $$
for some $\phi(\w,t)$.
Substituting for $d\y$, gives
$$ (\x-\y)^T\J\,d\w = d\phi, $$
and hence
$$ \y-\x = \Jinv \phi_\w, $$
which is of the form (\genfunc).} 

Now take the time derivative of (\genfunc), which yields
$$\left(\I-\half\Jinv\phi_{\w\w}\right)\dot\z=\Jinv\phi_{\w t}
\putnum$$\name\mmiddf
Let us now define a function $~H(\z,t)$ through
$$\phi_t(\w,t) = ~H(\z,t),\quad  \z=\w-\half\J\phi_\w,
\putnum$$\name\surrdef
the latter relation being a consequence of equation (\genfunc).
Taking the gradient of equation (\surrdef) and rearranging gives
$$\left(\I-\half\Jinv\phi_{\w\w}\right)\Jinv ~H_\z=\Jinv\phi_{\w t},
\putnum$$
and comparing this with equation (\mmiddf) gives
$$\dot\z=\Jinv ~H_\z(\z,t).  \putnum$$
In other words, the map (\genfunc) is equivalent to evolution under
the surrogate Hamiltonian $~H$. Equation (\surrdef) relating the
generating function and the surrogate Hamiltonian is a Hamilton-Jacobi
equation.

To derive the implicit midpoint method for a system with Hamiltonian $H$, we
set $\phi(\w,t)\equiv tH(\w)$; with this choice equation (\genfunc)
evaluated at $t=\tau$ is identical to equations (\fengform) except for minor
differences in notation. The surrogate Hamiltonian for the implicit midpoint
method with timestep $t$ 
is then defined implicitly by equation (\surrdef), which reads
$$~H\left[\w-\half t\J H_\w(\w)\right]=H(\w). \putnum$$\name\implic
We may write $~H=H+\Herr$, where the error Hamiltonian may be
expanded in a power series,
$$\Herr=\sum_{n=1}^\infty \Herr^nt^n.\putnum$$\name\herrdef
By substituting this expansion in (\implic) the functions $\Herr^1$,
$\Herr^2$, etc. can be determined in succession; for example, assuming
$H$ is autonomous we have 
$$\Herr^1=0,\qquad\Herr^2=\frac1/8\sum_{i,j,k,l=1}^{2K}
H_{i}J_{ij}H_{jk}J_{kl}H_{l},\putnum$$
where $H_i=\partial H/\partial w_i$, 
$H_{ij}=\partial^2 H/\partial w_i\partial w_j$. 
Of course, if the system is to be followed over more than one
timestep, the factor $t^n$ in equation (\herrdef) should be replaced
by $[t (\hbox{mod }\tau)]^n$.

For a nearly integrable Hamiltonian of the form (\sahag), the error
Hamiltonian becomes
$$\Herr= \frac1/8\epsilon t^2 \sum_{i,j=1}^K{\partial H^{(0)}\over
\partial p_i}{\partial^2 H^{(1)}\over\partial q_i\partial q_j}
{\partial H^{(0)}\over\partial p_j}+\hbox{O}(t^3).\putnum$$ 

\begingroup \parindent=0pt \frenchspacing
\everypar{\hangindent 20pt \hangafter 1}
 
\sectionlike References

Bellen, A., \& Zennaro, M. (1989), {\sl J. Comput. Appl. Math.,} {\bf
25}, 341

Feng, K. (1986), {\sl J. Comp. Math.,} {\bf 4}, 279

Kinoshita, H., Yoshida., H, \& Nakai, H. (1991), {\sl Cel. Mech. and
Dyn. Astr.,} {\bf 50}, 59

Laskar, J. (1989), {\sl Nature,} {\bf 338}, 237

Laskar, J. (1990), {\sl Icarus,} {\bf 88}, 266

Laskar, J. (1993), {\sl Physica D}, {\bf 67}, 257

Laskar, J. (1994), {\sl Astron. Astrophys.,} {\bf 287}, L9

Laskar, J., \& Roboutel, P. (1993), {\sl Nature}, {\bf 361}, 608

Ortega, J.M., \& Rheinbolt, W.C. (1970), {\sl Iterative solution of
nonlinear equations in several variables,} (Academic Press, New York)

Quinlan, G. D., \& Tremaine, S. (1990), {\sl Astron. J.,} {\bf 100},
1694

Quinn, T. R., \& Tremaine, S. (1990), {\sl Astron. J.,} {\bf 99}, 1016

Quinn, T. R., Tremaine, S., \& Duncan, M. (1991), {\sl Astron. J.,}
{\bf 101}, 2287

Saha, P., \& Tremaine, S. (1992), {\sl Astron. J.,} {\bf 104}, 1633

Saha, P., \& Tremaine, S. (1994), {\sl Astron. J.,} {\bf 108}, 1962 (\ST)

Sussman, G. J., \& Wisdom, J. (1992), {\sl Science,} {\bf 257}, 56

Touma, J., \& Wisdom, J. (1993), {\sl Science,} {\bf 259}, 1294

Wisdom, J., \& Holman, M. (1991), {\sl Astron. J.,} {\bf 102}, 1528
(\WH)

Wisdom, J., \& Holman, M. (1992), {\sl Astron. J.,} {\bf 104}, 2022

Wisdom, J., Holman, M., \& Touma, J. (1994), in {\it Integration Algorithms
for Classical Mechanics}, proceedings of a workshop at the Fields Institute
(\WHT)

 \endgroup

\closeout1 \vfill\eject
               \leftline{\bf\uppercase{figure captions}}
               \openout1=pscalls \input captions \closeout1
               \vfill\supereject \input pscalls \end